\documentclass[conference]{IEEEtran}
\IEEEoverridecommandlockouts
\usepackage{amsmath,amsfonts}
\usepackage[ruled,linesnumbered]{algorithm2e} 
\usepackage{array}
\usepackage[caption=false,font=normalsize,labelfont=sf,textfont=sf]{subfig}
\usepackage{textcomp}
\usepackage{stfloats}
\usepackage{url}
\usepackage{verbatim}
\usepackage{graphicx}
\usepackage{cite}
\usepackage{xcolor}
\usepackage{hyperref}
\usepackage{marvosym}
\usepackage{multirow}

\usepackage{booktabs}
\hypersetup{hidelinks}  

\usepackage{enumitem}
\usepackage{tabularray}
\usepackage{tikz}

\SetNlSty{textbf}{}{}  
\SetAlgoNlRelativeSize{-1}  
\DeclareMathOperator*{\argmin}{arg\,min}
\def\BibTeX{{\rm B\kern-.05em{\sc i\kern-.025em b}\kern-.08em
    T\kern-.1667em\lower.7ex\hbox{E}\kern-.125emX}}
    
\begin{document}
\title{Generative Semantic Communication for Text-to-Speech Synthesis}
\author{\IEEEauthorblockN{Jiahao Zheng$^{1,2,3*}$, Jinke Ren$^{1,2,3*}$, Peng Xu$^{1,2,3}$, Zhihao Yuan$^{1,2,3}$, \\Jie Xu$^{2,1,3}$, Fangxin Wang$^{2,1,3}$, Gui Gui$^{4}$, and Shuguang Cui$^{2,1,3}$} 
\IEEEauthorblockA{$^1$Shenzhen Future Network of Intelligence Institute,
$^2$School of Science and Engineering,\\
and $^3$Guangdong Provincial Key Laboratory of Future Networks of Intelligence,\\ The Chinese University of Hong Kong (Shenzhen), Shenzhen, China\\
$^4$School of Automation, Central South University, Changsha, China\\
E-mail: \{jiahaozheng1, pengxu1, zhihaoyuan\}@link.cuhk.edu.cn; \\ \{jinkeren, xujie, wangfangxin, shuguangcui\}@cuhk.edu.cn; guigui@csu.edu.cn} 
\thanks{The work was supported in part by NSFC with Grant No. 62293482, the Basic Research Project No. HZQB-KCZYZ-2021067 of Hetao Shenzhen-HK S\&T Cooperation Zone, the Shenzhen Outstanding Talents Training Fund 202002, the Guangdong Research Projects No. 2017ZT07X152 and No. 2019CX01X104, the Guangdong Provincial Key Laboratory of Future Networks of Intelligence (Grant No. 2022B1212010001), and the Shenzhen Key Laboratory of Big Data and Artificial Intelligence (Grant No. ZDSYS201707251409055). $*$ indicates equal contribution.~Jinke Ren is the corresponding author.}}
\maketitle

\begin{abstract}
Semantic communication is a promising technology to improve communication efficiency by transmitting only the semantic information of the source data. However, traditional semantic communication methods primarily focus on data reconstruction tasks, which may not be efficient for emerging generative tasks such as text-to-speech (TTS) synthesis. To address this limitation, this paper develops a novel generative semantic communication framework for TTS synthesis, leveraging generative artificial intelligence technologies. Firstly, we utilize a pre-trained large speech model called \textit{WavLM} and the residual vector quantization method to construct two semantic knowledge bases (KBs) at the transmitter and receiver, respectively. The KB at the transmitter enables effective semantic extraction, while the KB at the receiver facilitates lifelike speech synthesis. Then, we employ a transformer encoder and a diffusion model to achieve efficient semantic coding without introducing significant communication overhead. Finally, numerical results demonstrate that our framework achieves much higher fidelity for the generated speech than four baselines, in both cases with additive white Gaussian noise channel and Rayleigh fading channel.
\end{abstract}

\section{Introduction}\label{sec:intro}
The rapid proliferation of intelligent applications, such as augmented reality and the metaverse, presents significant challenges to next-generation wireless networks. Traditional communication technologies struggle to meet the massive data transmission requirements of these applications while ensuring ultra-reliability and low latency. To address these issues, semantic communication has emerged as a promising solution. By prioritizing the transmission of task-relevant information rather than the entire source data, semantic communication can significantly reduce communication overhead and enhance communication efficiency \cite{Lan}.

The appeal of semantic communication lies in its potential to extract and transmit semantic information from the source data. To date, numerous studies have explored semantic communication using deep learning algorithms \cite{Bourtsoulatze,Huang, Wu, Wang,Peng}. However, the majority of these works focus on data reconstruction tasks, such as image restoration \cite{Bourtsoulatze,Huang, Wu}  and text transmission \cite{Wang,Peng}. These designs often necessitate specialized models, thus limiting their generalizability across diverse contexts.

In recent years, the advent of generative artificial intelligence (GAI) technologies has motivated a growing interest in generative tasks, such as image-to-video generation and text-to-speech (TTS) synthesis. To meet the various requirements of generative tasks, a new generative semantic communication paradigm has been proposed, which aims to generate desired contents based on received signals and semantic knowledge bases (KB) \cite{Ren}. Specifically, the semantic KB is constructed by GAI technologies, which is capable of learning intrinsic data structures and generating new content not presented in the training data. Due to the significant advantages of semantic KBs, generative semantic communication offers a novel solution to undertake various tasks, such as image generation \cite{Ye} and remote surveillance \cite{Yang}.

In this paper, we consider a TTS synthesis task in a point-to-point communication system. The objective is to generate natural speech at the receiver, guided by a speech demonstration and a piece of text at the transmitter. To achieve this goal, there exist two common methods. One involves generating speech directly at the transmitter, and then transmitting the synthesized speech to the receiver. The other involves encoding the speech demonstration and text into digital signals for transmission, followed by generating speech at the receiver based on the received signals. However, both methods encounter high communication overhead. To improve the communication efficiency, a pioneering work \cite{Qin} has proposed a deep learning-based semantic communication system for speech recognition and synthesis. However, the resultant computational cost at the receiver is intensive due to the employment of complex speech synthesis algorithms.

To reduce both computational costs and communication overhead, this paper proposes to incorporate two semantic KBs to enable semantic extraction and speech synthesis at the transmitter and receiver, respectively. Specifically, the two semantic KBs are constructed using a pre-trained large speech model called WavLM \cite{Chen}, which is able to extract the key speech characteristics from the speech demonstration. Additionally, a transformer encoder is employed to extract the residual information from the speech demonstration that is not captured by the KB at the transmitter. Furthermore, a diffusion-based semantic decoder is developed to generate lifelike speech using the KB at the receiver. Simulation results show that our method produces higher-quality speech and is more resilient to channel dynamics compared to traditional syntactic communication and existing semantic communication methods.

The rest of this paper is organized as follows. Section II introduces the system model. Section III presents the specific designs of the semantic KBs and semantic decoder. Section IV provides a two-stage training algorithm. Section V discusses the experimental results and Section VI concludes the paper.
\begin{figure*}[htbp]
\centering
\includegraphics[width=16.2cm]{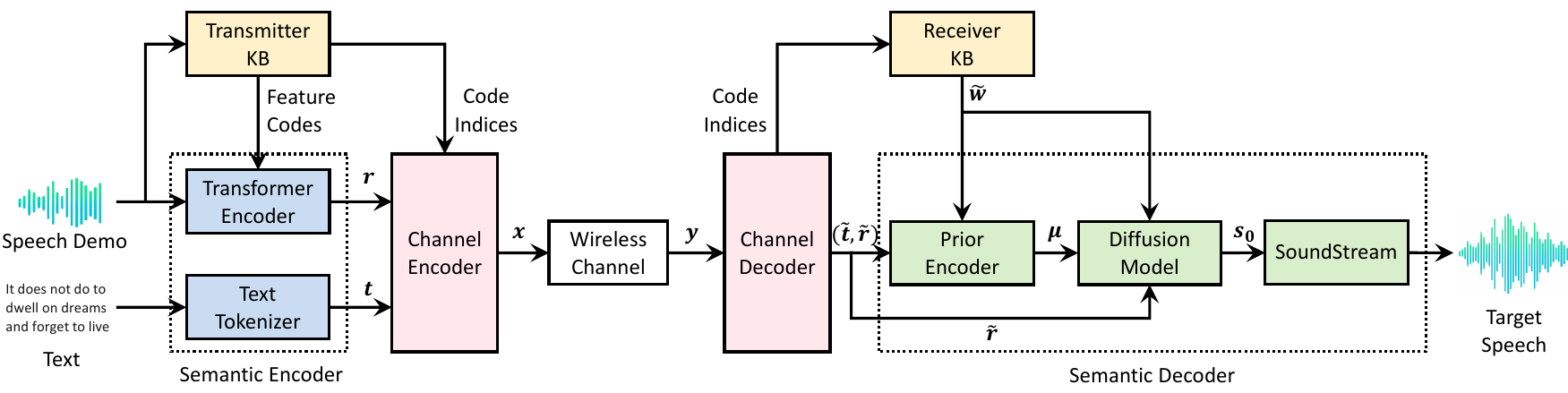}
\caption{Generative semantic communication system for text-to-speech synthesis.}\label{fig: system}
\end{figure*}

\section{System Model}\label{sec:methodology}
As shown in Fig. \ref{fig: system}, we consider a point-to-point communication system, in which a transmitter possesses a piece of text and a speech demonstration, while a receiver aims to generate a target speech that aligns with both the content of the text and the characteristics of the speech demonstration. Instead of encoding the speech demonstration and text into bit sequences for transmission, we consider the semantic communication technology. Specifically, two semantic KBs are deployed at the transmitter and receiver to facilitate semantic extraction and speech synthesis, respectively. The detailed working mechanism is described as follows. At the transmitter, a text tokenizer converts the input text into a token vector $\boldsymbol{t} \in \mathbb{R}^{d_t}$, where ${d_t}$ is the dimension of the token vector. Meanwhile, the speech demonstration is input into the transmitter KB, which outputs a set of feature codes as well as the corresponding code indices. The feature codes contain the key knowledge necessitated for speech synthesis, which are input into a transformer encoder. Then, the transformer encoder extracts the residual information $\boldsymbol{r} \in \mathbb{R}^{d_r}$ from the speech demonstration not captured by the feature codes, where $d_r$ denotes the dimension of the residual vector. Notably, the combination of the transformer encoder and the text tokenizer is defined as the semantic encoder.

After semantic encoding, the token vector $\boldsymbol{t}$, residual vector $\boldsymbol{r}$, and the code indices are embedded into one packet, which is encoded by a channel encoder before transmitting to the receiver. In this work, we use a neural network to perform channel encoding. It consists of two cascaded modules, in which each module includes one convolutional layer followed by a transformer encoder. Let $\boldsymbol{x} \in \mathbb{R}^{d_x}$ denote the transmit signal, where $d_x$ is the dimension of the output of channel encoder. Then, the received signal is expressed as  
\begin{equation}\label{eql:channel}
\boldsymbol{y} = h \boldsymbol{x} + \boldsymbol{n},
\end{equation}
where $h$ denotes the channel coefficient, $\boldsymbol{n} \sim  \mathcal{CN}(0, \sigma^2 \boldsymbol{I})$ represents the independent and identically distributed circularly symmetric complex Gaussian noise vector with noise power $\sigma^2$, and $\boldsymbol{I}$ is an identity matrix. 

Once receiving the signal $\boldsymbol{y}$, the receiver performs channel decoding to obtain the decoded token vector $\tilde{\boldsymbol{t}} \in \mathbb{R}^{d_t}$, residual vector $\tilde{\boldsymbol{r}} \in \mathbb{R}^{d_r}$, and code indices. Similar to the channel encoder, the channel decoder also consists of two cascaded modules, where each module includes a trans-convolutional layer and a transformer encoder. Subsequently, the code indices are input into the receiver KB to reconstruct the speech feature corresponding to the speech demonstration, denoted by $\tilde{\boldsymbol{w}}$ with dimension ${d_w}$.
Finally, $\tilde{\boldsymbol{w}}$, $\tilde{\boldsymbol{t}}$, and $\tilde{\boldsymbol{r}}$ are all input into a semantic decoder to generate the target speech. 

According to the above working mechanism, we observe that the two semantic KBs and the semantic decoder are the key components of the whole system. Therefore, we present their specific designs in the next section.
\begin{figure} 
\centering
\includegraphics[width=7.5cm]{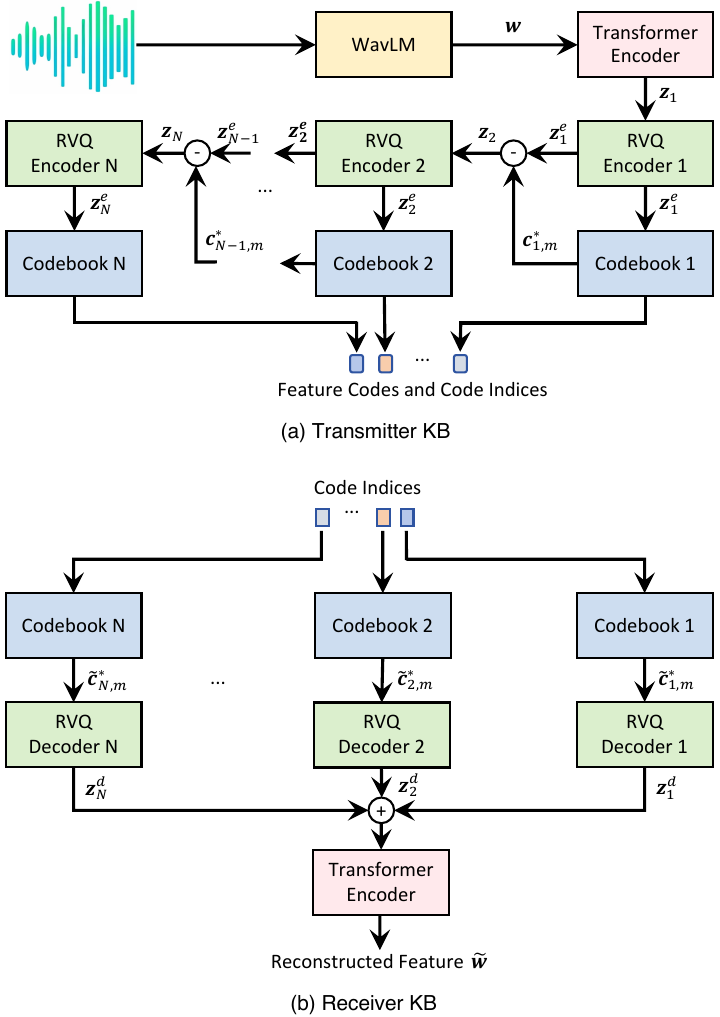}
\caption{Illustration of the semantic KBs.}
\label{fig: KB}
\end{figure}
\begin{figure} 
\centering
\includegraphics[width=7.5cm]{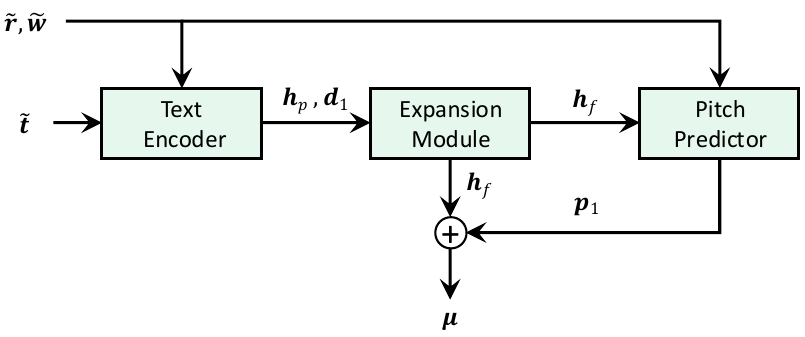}
\caption{Illustration of the prior encoder.}
\label{fig: prior}
\end{figure}

\section{Semantic KBs and Semantic Decoder}
\subsection{Design of Semantic KBs}
Fig. \ref{fig: KB}(a) shows the structure of the semantic KB at the transmitter. It consists of a large speech model called WavLM \cite{Chen}, a transformer encoder, $N$ residual vector quantization (RVQ) encoders \cite{Defossez}, and $N$ feature codebooks.

WavLM is a pre-trained large speech model, which can extract key information from speeches and has been widely employed to solve full-stack downstream speech tasks. Therefore, it serves as the key component of the transmitter KB. Specifically, WavLM extracts a speech feature vector from the the speech demonstration, denoted by $\boldsymbol{w} \in \mathbb{R}^{d_w}$ with dimension $d_w$. For convenience, we name $\boldsymbol{w}$ as the WavLM feature, which reflects the characteristics of the speech demonstration, such as pitch, timbre, and loudness. 

Since the dimension of $\boldsymbol{w}$ is generally large, we compress it to improve communication efficiency. As the speech feature $\boldsymbol{w}$ is compositional hierarchies, we propose a multi-scale hierarchical model to compress it. As illustrated in Fig. \ref{fig: KB}(a), a transformer encoder first  converts $\boldsymbol{w}$ into a source vector $\boldsymbol{z}_1 \in \mathbb{R}^{d_w}$. Next, $\boldsymbol{z}_1$ is sequentially quantized by $N$ RVQ encoders, where each RVQ encoder extracts the remained speech information from the output of the previous RVQ encoder. Let $\boldsymbol{z}_{i}$ denote the input of the $i$-th RVQ encoder. Then, its output is defined as
\begin{equation}\label{eql:RVQ_output}
\boldsymbol{z}_{i}^{e}= \mathsf{RVQE}(\boldsymbol{z}_{i}; \boldsymbol{\theta}_{\text{RVQE},i}),
\end{equation}
where $\boldsymbol{\theta}_{\text{RVQE},i}$ is the parameter of the $i$-th RVQ encoder. Moreover, each RVQ encoder connects with a particular feature codebook, which is composed of $M$ feature codes for speech representation. Let $\boldsymbol{C}_{{i}} = [\boldsymbol{c}_{{i,1}},\boldsymbol{c}_{{i,2}},...,\boldsymbol{c}_{{i,M}}]$ denote the feature codebook associated with the $i$-th RVQ encoder, where $\boldsymbol{c}_{{i,m}}$ represents the $m$-th feature code. Next, the output of each RVQ encoder $\boldsymbol{z}_{i}^{e}$ is mapped to a feature code $\boldsymbol{c}^*_{{i,m}}$ via the nearest neighbor look-up method, which is given by
\begin{equation} \label{eql:mapping}
\boldsymbol{c}^*_{{i,m}} = \argmin_{\boldsymbol{c}_{{i,m}}}{\Vert \boldsymbol{c}_{{i,m}} - \boldsymbol{z}_{i}^{e} \Vert}. 
\end{equation}
Then, the input of the $(i+1)$-th RVQ encoder is computed by
\begin{equation}\label{eql:RVQ_input}
\boldsymbol{z}_{i+1} = \boldsymbol{z}_{i}^{e} - \boldsymbol{c}^*_{{i,m}}.
\end{equation}

We note that $\boldsymbol{z}_{i+1}$ contains the fine-grained speech information of $\boldsymbol{z}_{i}^{e}$ not captured by $\boldsymbol{c}^*_{{i,m}}$. Based on this hierarchical compression model, the WavLM feature $\boldsymbol{w}$ is finally represented by $N$ feature codes, i.e., $\{\boldsymbol{c}^*_{{1,m}},\boldsymbol{c}^*_{{2,m}},\cdots,\boldsymbol{c}^*_{{N,m}}\}$.

As shown in Fig. \ref{fig: KB}(b), the KB at the receiver is composed of $N$ feature codebooks, $N$ RVQ decoders, and a transformer encoder. The feature codebooks in the receiver KB are the same as those in the transmitter KB. Therefore, we use the code indices to search the feature codes output by the transmitter KB. Let $\widetilde{\boldsymbol{c}}^*_{{i,m}}$ denote the feature code that matches the $i$-th code index. Then, each $\widetilde{\boldsymbol{c}}^*_{{i,m}}$ is further input into a RVQ decoder to compute a latent feature
\begin{equation}\label{eql:latent_feature}
    \boldsymbol{z}_{i}^{d}= \mathsf{RVQD}(\widetilde{\boldsymbol{c}}^*_{{i,m}}; \boldsymbol{\theta}_{\text{RVQD},i}),
\end{equation}
where $\boldsymbol{\theta}_{\text{RVQD},i}$ is the parameter of the $i$-th RVQ decoder. Finally, the outputs of all RVQ decoders are summed up, and the result is input into a transformer encoder to reconstruct the WavLM feature, denoted by $\tilde{\boldsymbol{w}} \in \mathbb{R}^{d_w}$.

\subsection{Design of Semantic Decoder}
As shown in the right half of Fig. \ref{fig: system}, the semantic decoder is composed of three modules, including a prior encoder \cite{Shen}, a diffusion model \cite{Popov}, and a SoundStream \cite{Defossez}.

The prior encoder is designed to generate conditional information, which controls the speech synthesis process of the diffusion model. As illustrated in Fig. \ref{fig: prior}, the prior encoder consists of a text encoder, an expansion module, and a pitch predictor. The text encoder and pitch predictor adopt the transformer encoder structure. Specifically, by taking the token vector $\tilde{\boldsymbol{t}}$, reconstructed WavLM feature $\tilde{\boldsymbol{w}}$, and residual vector $\tilde{\boldsymbol{r}}$ as input, the text encoder outputs a phoneme-level hidden sequence $\boldsymbol{h}_p$ and a duration vector $\boldsymbol{d}_1$ of the phonemes. Next, $\boldsymbol{h}_p$ is expanded to a frame-level hidden sequence $\boldsymbol{h}_f$ according to the duration vector $\boldsymbol{d}_1$. Then, the pitch predictor uses $\tilde{\boldsymbol{r}}$, $\tilde{\boldsymbol{w}}$, and $\boldsymbol{h}_f$ to predict the pitch information $\boldsymbol{p}_1$ of the target speech. Finally, $\boldsymbol{p}_1$ and $\boldsymbol{h}_f$ are summed up to obtain the final conditional information, denoted by $\boldsymbol{\mu}$.

The diffusion model aims to generate the SoundStream feature associated with the target speech. Its working mechanism includes a forward process and a backward process, which can be formulated as two stochastic differential equations. Specifically, the forward process adds noise to a ground-truth SoundStream feature $\boldsymbol{s}^{\text{g}}_0$, transforming it into Gaussian noise $\boldsymbol{s}_T$, where $T \in [0,1]$ is the normalized step. The backward process removes the noise from $\boldsymbol{s}_T$ and predicts the SoundStream feature, denoted by $\boldsymbol{s}_0$. According to  \cite{Popov}, the forward process is expressed as
\begin{equation}\label{eql:forward}
\boldsymbol{s}_T = e^{-\frac{1}{2}\int_{0}^{T}{\beta_tdt}} \boldsymbol{s}^{\text{g}}_0 + \int_{0}^{T}{\sqrt{\beta_s}e^{-\frac{1}{2}\int_{0}^{T}{\beta_t}dt}d\boldsymbol{\omega}_s},
\end{equation}
where $\boldsymbol{\omega}_s$ denotes standard Brownian motion, while $\beta_t$ and $\beta_s$ are two schedule functions for noise prediction.

In the backward process, for each normalized step $t \in [0, T]$, the noisy SoundStream feature $\boldsymbol{s}_t$ given $\boldsymbol{s}_0$ follows a Gaussian distribution, i.e., $p(\boldsymbol{s}_t\mid{\boldsymbol{s}_0}) \sim \mathcal{N}(\boldsymbol{\rho}_t(\boldsymbol{s}_0), \sigma_t \boldsymbol{I})$, where $\boldsymbol{\rho}_t(\boldsymbol{s}_0)$ is the mean vector,  $\sigma_t$ is the variance value, and $\boldsymbol{I}$ is an identity matrix \cite{Popov}. Let $\nabla{\log {p_t(\boldsymbol{s}_t)}}$ denote the gradient vector of the log probability density function of $\boldsymbol{s}_t$. Then, the backward process for predicting the SoundStream feature $\boldsymbol{s}_0$ is given by
\begin{equation}\label{eql:backward}
\boldsymbol{s}_0 = \boldsymbol{s}_T + \int_{0}^{T}{-\left(\frac{1}{2} \boldsymbol{s}_t + \nabla{\log{p_t(\boldsymbol{s}_t)}}\right)\beta_tdt}.
\end{equation}

We note that both the forward and backward processes are performed during the training stage, but only the backward process operates during the inference stage. Specifically, for each normalized step $t \in [0, T]$ in the inference stage, $\boldsymbol{\mu}$, $\tilde{\boldsymbol{w}}$, $\tilde{\boldsymbol{r}}$ and $\boldsymbol{s}_t$ are all input into a WaveNet \cite{Oord} to predict the value of $\boldsymbol{\rho}_t(\boldsymbol{s}_0)$. Then, we approximate the probability density function $p_t(\boldsymbol{s}_t)$, followed by computing the SoundStream feature $\boldsymbol{s}_0$ according to \eqref{eql:backward}. Finally, the SoundStream converts the SoundStream feature $\boldsymbol{s}_0$ to the target speech.

\begin{algorithm}[t] 
\caption{Training Algorithm in the Second Stage.}\label{Alg.1}
    \KwIn{Speech demonstration $\boldsymbol{D}$, original text $\boldsymbol{T}$, number of training epochs $J$.}
    
     Initialize parameters $\boldsymbol{\theta}_{r}^0, \boldsymbol{\theta}_{e}^0,\boldsymbol{\theta}_{d}^0,\boldsymbol{\theta}_{p}^0, \boldsymbol{\theta}_{\text{diff}}^0$.\\    
     $\boldsymbol{t} \leftarrow \mathtt{Text~Tokenizer}(\boldsymbol{T})$.\\
     Feature codes and indices $\leftarrow \mathtt{Transmitter~KB}(\boldsymbol{D})$.\\
     \For {$j= 0, 1, \cdots, J-1$}
     {
         $\boldsymbol{r}^j \!\!\leftarrow \!\!\mathtt{Transformer~Encoder}({\text{feature codes}},\boldsymbol{D};\boldsymbol{\theta}_{r}^{j})$.\\
         $\boldsymbol{f}_e^j \leftarrow \mathtt{Padding}({\text{code indices}},\boldsymbol{r}^j, \boldsymbol{t})$.\\
         $\boldsymbol{x}^j \leftarrow \mathtt{Channel~Encoder}(\boldsymbol{f}_e^j;\boldsymbol{\theta}_{e}^j)$.\\
         Transmit $\boldsymbol{x}^j$ to the receiver via $\boldsymbol{y}^j = h^j {\boldsymbol{x}}^j + {\boldsymbol{n}}^j$.\\
         $\boldsymbol{f}_d^j \leftarrow \mathtt{Channel~Decoder}(\boldsymbol{y}^j;\boldsymbol{\theta}_{d}^{j})$.\\
         Identify $\boldsymbol{\widetilde{r}}^j, \boldsymbol{\widetilde{t}}^j,$ and code indices from $\boldsymbol{f}_d^j$.\\       
         $\widetilde{\boldsymbol{w}}^j \leftarrow \mathtt{Receiver~KB}({\text{code indices}})$.\\
         $\boldsymbol{\mu}^j \leftarrow \mathtt{Prior~Encoder}(\widetilde{\boldsymbol{r}}^j,\widetilde{\boldsymbol{t}}^j,\widetilde{\boldsymbol{w}}^j; \boldsymbol{\theta}_{p}^j)$.\\
         $\boldsymbol{s}_0^j \leftarrow \mathtt{Diffusion~Model}(\boldsymbol{\mu}^j, \widetilde{\boldsymbol{w}}^j, \widetilde{\boldsymbol{r}}^j; \boldsymbol{\theta}_{\text{diff}}^j)$.\\ 
         Compute the loss function $L_{\text{total}}$ according to \eqref{eql:total_loss}.\\
         Perform stochastic gradient descent to minimize  $L_{\text{total}}$ and obtain $\boldsymbol{\theta}_{r}^{j+1}$, $\boldsymbol{\theta}_{e}^{j+1}$, $\boldsymbol{\theta}_{d}^{j+1}$, $\boldsymbol{\theta}_{p}^{j+1}$, $\boldsymbol{\theta}_{\text{diff}}^{j+1}$.
     }
    \KwOut{Trained models $\boldsymbol{\theta}_{r}^J$,  $\boldsymbol{\theta}_{e}^J$, $\boldsymbol{\theta}_{d}^J$, $\boldsymbol{\theta}_{p}^J$, $\boldsymbol{\theta}_{\text{diff}}^J$.}
\end{algorithm}

\section{Two-stage Training Algorithm}
This section presents a two-stage training algorithm for the proposed generative semantic communication system. To mitigate the influence of other modules on the semantic KBs, we first train the semantic KBs before proceeding to train other modules, thus enhancing the reliability of the semantic KBs.

In the first stage, all modules except the two semantic KBs are fixed and the channel impairments are ignored. Moreover, we adopt a pre-trained WavLM model in \cite{Chen} and fix it across the training process. 
Define $\boldsymbol{\Theta}_{{\text{RVQE}}}=\left[\boldsymbol{\theta}_{{\text{RVQE},1}}, \cdots, \boldsymbol{\theta}_{{\text{RVQE},N}}\right], \boldsymbol{\Theta}_{{\text{RVQD}}}=\left[\boldsymbol{\theta}_{{\text{RVQD},1}},\cdots, \boldsymbol{\theta}_{{\text{RVQD},N}}\right]$, $\boldsymbol{\phi}_{{t}}$, and $\boldsymbol{\phi}_{{r}}$ as the parameters of all RVQ encoders, RVQ decoders, the transformer encoder in the transmitter KB, and the transformer encoder in the receiver KB, respectively. Then, the loss function for training semantic KBs can be divided into three parts: 1) reconstruction loss, which is defined as the mean squared error of the WavLM feature $\boldsymbol{w}$ and the reconstructed WavLM feature $\tilde{\boldsymbol{w}}$, i.e., $L_r (\boldsymbol{\Theta}_{{\text{RVQE}}},\boldsymbol{\Theta}_{{\text{RVQD}}}, \boldsymbol{\phi}_{{t}}, \boldsymbol{\phi}_{{r}}, \boldsymbol{C}_i) = \Vert \boldsymbol{w} - \tilde{\boldsymbol{w}} \Vert_2^2$,  where $\Vert\cdot\Vert_2$ denotes the L2 norm; 2) embedding loss for all feature codebooks, which is defined as  $L_e (\boldsymbol{C}_i) = \sum_{i=1}^{N} \Vert \mathsf{ng}[{\boldsymbol{z}}_{i}^{e}] - 
\widetilde{\boldsymbol{c}}_{i,m}^* \Vert_2^2$, where $\mathsf{ng}[\cdot]$ is the non-gradient operator, indicating that there is no gradient passed to the variable and its gradients is zero \cite{Hu}; and 3) commitment loss for all RVQ encoders, i.e., $L_c (\boldsymbol{\Theta}_{\text{RVQE}}, \boldsymbol{\phi}_{{t}}) = \sum_{i=1}^{N} \Vert \boldsymbol{z}_{i}^{e} - \mathsf{ng}[\widetilde{\boldsymbol{c}}_{i,m}^*] \Vert_2^2$. Hence, the training loss function for the two semantic KBs is given by 
\begin{equation}\label{eql:KB_loss}
\begin{aligned}
&L_{\text{KB}}(\boldsymbol{\Theta}_{{\text{RVQE}}},\boldsymbol{\Theta}_{{\text{RVQD}}}, \boldsymbol{\phi}_{{t}},\boldsymbol{\phi}_{{r}},\boldsymbol{C}_i)\\
&=\alpha_1 \times L_r (\boldsymbol{\Theta}_{{\text{RVQE}}},\boldsymbol{\Theta}_{{\text{RVQD}}}, \boldsymbol{\phi}_{{t}},\boldsymbol{\phi}_{{r}},\boldsymbol{C}_i)\\&~~~~+\alpha_2 \times L_e (\boldsymbol{C}_i)+\alpha_3 \times L_c (\boldsymbol{\Theta}_{\text{RVQE}}, \boldsymbol{\phi}_{{t}}),
\end{aligned}
\end{equation}
where $\alpha_1$, $\alpha_2$, and $\alpha_3$ are weight coefficients.

In the second stage, we train other modules based on the well-trained semantic KBs. At the transmitter, the token vector $\boldsymbol{t}$, residual vector $\boldsymbol{r}$, and code indices are padded to the same dimension and concatenated into a whole vector, denoted by $\boldsymbol{f}_e$. Next, $\boldsymbol{f}_e$ is passed through the channel encoder, wireless channel, and channel decoder. Let $\boldsymbol{f}_d$ denote the output vector of the channel decoder. Then, the training loss function for the channel encoder and channel decoder is expressed as
\begin{equation}\label{eql:ED_loss}
L_{\text{ED}}(\boldsymbol{\theta}_{e},\boldsymbol{\theta}_{d}) = \Vert\boldsymbol{f}_e - \boldsymbol{f}_d \Vert,
\end{equation}
where $\boldsymbol{\theta}_{e}$ and $\boldsymbol{\theta}_{d}$ are the parameters of the channel encoder and channel decoder, and $\Vert \cdot \Vert$ represents the Euclidean norm.

Let $\boldsymbol{\theta}_{p}$ and $\boldsymbol{\theta}_{r}$ denote the parameters of the prior encoder and the transformer encoder in the semantic encoder. Then, the training loss function for the prior encoder and transformer encoder is given by
\begin{equation}\label{eql:PT_loss}
L(\boldsymbol{\theta}_{p}, \boldsymbol{\theta}_{r}) = \Vert \boldsymbol{d}_1 - \boldsymbol{d}_0 \Vert + \Vert \boldsymbol{p}_1 -\boldsymbol{p}_0 \Vert + \Vert \boldsymbol{\mu} - \boldsymbol{s}_0 \Vert,
\end{equation}
where $\boldsymbol{d}_0$ and $\boldsymbol{p}_0$ are the ground-truth phoneme-level duration vector and the frame-level pitch information of the training sample, respectively. Furthermore, let $\boldsymbol{\theta}_{\text{diff}}$ denote the parameter of the diffusion model. Then, for a given normalized step $t$, the training loss function for the diffusion model and transformer encoder is defined as 
\begin{equation}\label{eql:diffusion_loss}
L(\boldsymbol{\theta}_{\text{diff}},\boldsymbol{\theta}_{r}) = \Vert \nabla{\log{p_t(\boldsymbol{s}_t)}} - \sigma^{-1}_{t}(\boldsymbol{\rho}_t(\boldsymbol{s}_0)-\boldsymbol{s}_t)\Vert.
\end{equation}

Based on the above analysis, the total loss function in the second stage is given by
\begin{equation}\label{eql:total_loss}
L_{\text{total}} = L_{\text{ED}}(\boldsymbol{\theta}_{e},\boldsymbol{\theta}_{d}) + L(\boldsymbol{\theta}_{p}, \boldsymbol{\theta}_{r}) + L(\boldsymbol{\theta}_{\text{diff}},\boldsymbol{\theta}_{r}).
\end{equation}
The corresponding training algorithm is given in Algorithm \ref{Alg.1}. 

\section{Experimental Results}
In this section, we conduct experiments to demonstrate the effectiveness of the proposed framework. All experiments are implemented in a Linux environment equipped with one NVIDIA Tesla A100 40GB GPU.
\subsection{Experiment Settings}\label{sec:settings}
\textbf{Datasets.} We use the LibriTTS dataset \cite{Zen} for experiments, which is a multi-speaker English corpus of nearly 585 hours of read English speech at a 24kHz sampling rate. Our model is trained on the LibriTTS train-other set and is evaluated on the LibriTTS test-other set.

\textbf{Models.} In the transmitter KB, we use a pre-trained WavLM feature extractor, i.e., the module before the post-project layer \cite{Chen}, to reduce the computational overhead at the transmitter. The numbers of RVQ encoders and codebooks are both set as $N=8$. Specifically, each RVQ encoder contains a linear projection layer followed by a residual block, and each RVQ decoder contains a residual block followed by a linear projection layer. The number of feature codes of each feature codebook is set as $M=1024$. The weight coefficients in \eqref{eql:KB_loss} are set as $\alpha_1=\alpha_2 =1$ and $\alpha_3=0.25$. The diffusion model contains 24 WaveNet layers, each consisting of 1D dilated convolution layer. Other configurations are shown in Table~\ref{tab: model config}.
\begin{table}[t]
  \caption{Model Configurations}
  \vspace{-.5em}
  \label{tab: model config}
  \centering
  \begin{tabular}{ ccc }
    \toprule
    \textbf{Module} & \textbf{Configuration} & \textbf{Value}  \\
    \midrule
    \multirow{7}{*}{Prior encoder} 
                & Conv1D layers  &   30\\
                & Conv1D kernel size    & 5   \\
                & Conv1D filter size & 1024 \\
                & Attention layers       & 18   \\
                & Attention heads           & 4   \\
                & Hidden dimension          & 256   \\
                & Dropout                   & 0.2  \\
    \midrule
                & Attention layers          & 3   \\
    {Transformer encoder}         & Attention heads           & 4   \\
                
    (in semantic encoder)            & Hidden dimension          & 256  \\
                & Dropout                   & 0.2  \\
    \midrule
    \multirow{4}{*}{Channel encoder} & Conv1D layers  &  4\\
                 & Attention layers          & 2   \\
                & Attention heads           & 4   \\
                & Hidden dimension          & 256   \\
    \midrule
    \multirow{4}{*}{Channel decoder} & Conv1D layers  & 4\\
                & Attention layers          & 2   \\
                & Attention heads           & 4   \\
                & Hidden dimension          & 256   \\
    \midrule
    \multirow{3}{*}{Diffusion model} &WaveNet layers  &  24 \\
                & Hidden dimension          & 256  \\
                & Dropout                   & 0.2  \\
    \bottomrule
  \end{tabular}
  \vspace{-1em}
\end{table}

\textbf{Baselines.} We consider four baseline schemes for performance comparison: 1) PCM+LDPC+YourTTS, which uses the classical pulse-code modulation (PCM) method and low-density parity-check (LDPC) code for separate source and channel coding, followed by a popular YourTTS \cite{Casanova} algorithm to generate the target speech; 2) PCM+LDPC+NS2, which uses the PCM method and LDPC code for separate source and channel coding, while using the state-of-the-art NaturalSpeech2 (NS2) algorithm \cite{Shen} for speech synthesis; 3) JSCC+YourTTS, which performs joint semantic and channel coding (JSCC) \cite{Bourtsoulatze} and uses the YourTTS algorithm for speech synthesis; 4) JSCC+NS2, which employs the JSCC method \cite{Bourtsoulatze} while using the NS2 algorithm to generate the target speech. In the NS2 algorithm, the speech demonstration is converted to an audio codec latent vector before being input into the channel encoder. The main difference between the four baselines and our scheme is that we incorporate two semantic KBs.

\textbf{Metrics.}
We consider two performance metrics to evaluate the quality of the synthesized speech. The first one is word error rate (WER), which measures the accuracy of the synthesized speech with respect to the original text. Lower WER indicates better speech. The second one is speaker similarity score (SPK), which represents the degree of similarity between the vocal characteristics of the synthesized speech and the ground-truth speech. The higher the SPK is, the higher probability that the speech pairs come from the same speaker. 

\begin{figure}[t]
\centering
\includegraphics[width=0.98\linewidth]{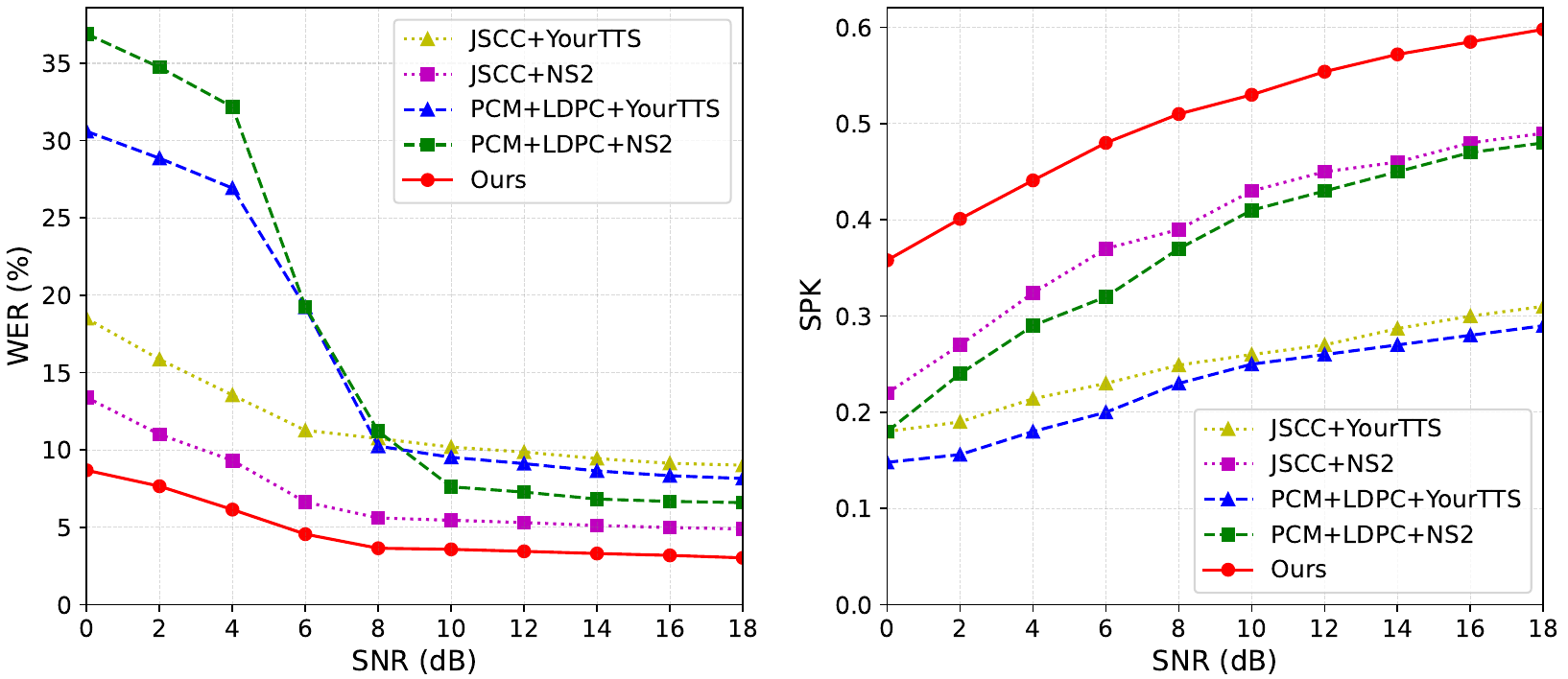}
\caption{WER and SPK of different schemes with AWGN channel.}\label{fig: AWGN}
\end{figure}
\begin{figure}[t]
\centering
\includegraphics[width=0.98\linewidth]{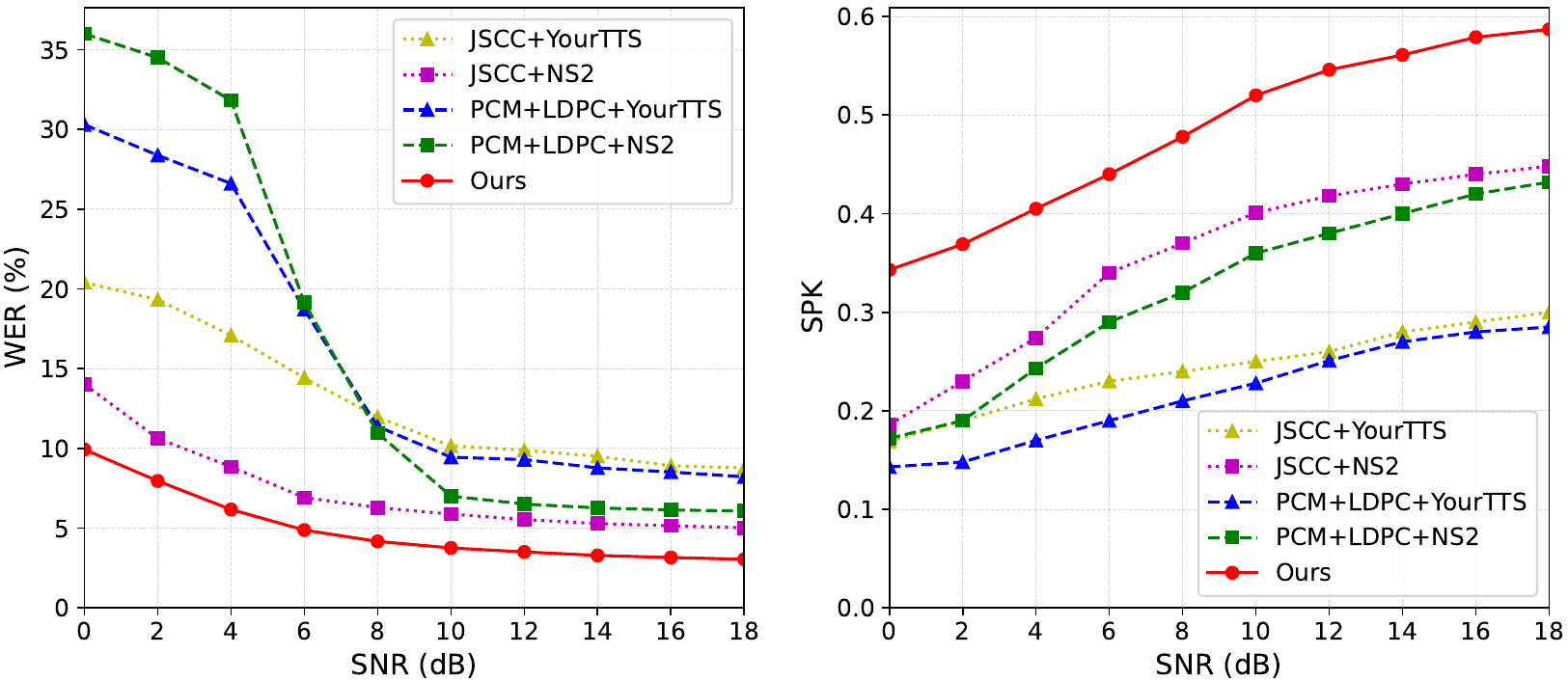}
\caption{WER and SPK of different schemes with Rayleigh fading channel.}\label{fig: Rayleigh}
\end{figure}
\subsection{Performance Comparison with Different Channels}\label{sec: different noise}
We first compare the performance of our generative scheme with the four baseline schemes. Both additive white Gaussian noise (AWGN) channel and Rayleigh fading channel are simulated. The data sizes of the original text and the speech demonstration are 6.4 Kbits and 2,304 Kbits, respectively. For fair comparison, the communication overhead of each scheme is bounded by 160 Kbits. The results are shown in Fig. \ref{fig: AWGN} and Fig. \ref{fig: Rayleigh}, respectively. It is observed that our generative scheme always achieves the lowest WER and highest SPK among all schemes in both AWGN and Rayleigh fading scenarios, demonstrating the effectiveness and generalization of our generative scheme. Moreover, in low signal-to-noise ratio (SNR) regimes, the two baselines associated with JSCC achieve lower WER and higher SPK than the other two baselines associated with PCM and LDPC, reflecting the superiority of semantic communication over traditional communication.

\begin{table}[t]
\footnotesize
\centering
\caption{Computation and Storage of Different Schemes}
\vspace{-.5em}
\renewcommand\arraystretch{1.1}
    \begin{tabular}{lccc}
        \toprule
        \multirow{2}{*}{Schemes}  & Computation             & Storage\\ 
                                  & ($\times10^{10}$ FLOPs) & (MByte)\\
        \midrule
        PCM+LDPC+YourTTS          & 21                      & 406\\
        PCM+LDPC+NS2              & 15213                   & 1908\\
        JSCC+YourTTS              & 29                      & 425\\
        JSCC+NS2                  & 15222                   &1919\\
        \textbf{Ours}             &\textbf{3878}             &\textbf{873}\\
        \bottomrule
    \end{tabular}
    \label{tab:comp_stog}
\end{table}

We further evaluate the computation and storage requirements for all schemes, and the results are presented in Table~\ref{tab:comp_stog}. It is observed that our generative scheme requires fewer computations and less storage than the two baseline schemes associated with NS2. However, the two baseline schemes associated with YourTTS demand even fewer computations and less storage than our scheme. Despite this advantage, the synthesized speech quality of the YourTTS baseline schemes is unsatisfactory. Therefore, our scheme achieves a better balance between computation, storage, and speech quality.

\subsection{Performance Comparison with Different Communication Budgets}\label{sec: communication overhead}
\begin{figure}[t]
\centering
\includegraphics[width=0.99\linewidth]{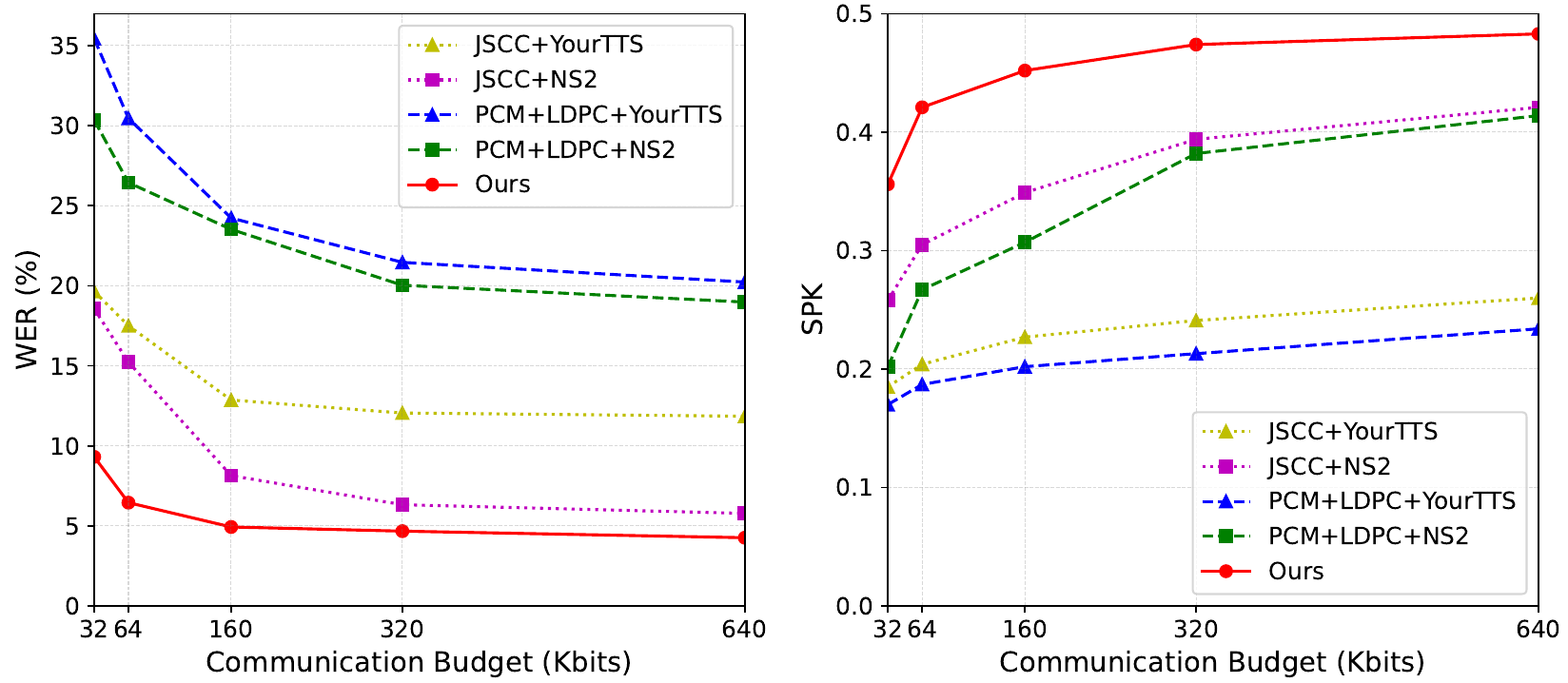}
\caption{WER and SPK of all schemes with different communication budgets.}
\label{fig:communication budget}
\end{figure}

To validate the robustness of the proposed generative scheme, we test the performance of all schemes with different communication budgets, i.e., the number of bits being transmitted, which determines the communication latency. We take the AWGN channel as an example and set SNR as 5dB. The result is shown in Fig. \ref{fig:communication budget}. It is seen that our generative scheme outperforms the four baselines. In particular, when the communication budget decreases, the performance gaps between the proposed scheme and other four schemes become more evident. This demonstrates that our generative scheme can generate better speech in communication-limited scenarios.

\section{Conclusion}\label{sec:conclusion}
This paper proposed a new generative semantic communication framework for TTS synthesis. Two semantic KBs are constructed leveraging the large speech model WavLM, which is able to capture the key characteristics of the speech demonstration. By integrating the two semantic KBs into the transmitter and receiver, along with the utilization of a transformer encoder and a diffusion model for semantic coding, our framework generates lifelike speech with low communication overhead and computational cost. This work opens a new direction for future research on semantic communication, highlighting the promising applications of GAI technologies in various communication-intensive tasks.

\end{document}